\newcommand{\AQO}[1]{\bar#1\gamma_3\gamma_5#1}
\newcommand{\PME}[1]{\langle\,p^\uparrow\,|\;#1\;|\,p^\uparrow\,\rangle}
\newcommand{\M}{\hphantom{-}}
\newcommand{\Z}{\hphantom{0}}
\def\abbreviation#1#2#3{\def#1{#3 (#2)\def#1{#2}}}
\abbreviation {\BSR}   {BSR}  {Bjorken sum rule}
\abbreviation {\DIS}   {DIS}  {deep-inelastic scattering}
\abbreviation {\PQCD}  {PQCD} {perturbative QCD}
\abbreviation {\QPM}   {QPM}  {quark-parton model}
\documentstyle[12pt,epsf,world_sci]{article}
\let\thecase=\relax
\def\topline#1{\raisebox{48pt}[0pt][0pt]{\makebox[\textwidth]{#1}}\relax}
\begin{document}
\title{\topline{July 1994 \hfill MITH 94/10}
       \thecase{Higher Twist, Polarised Structure Functions}\\
       \thecase{and the Bjorken System of Equations}
       \thanks{~Presented at the 9th Winter Course on Hadronic Physics,
                Folgaria, Italy, February 1994}
}
\author{Philip G. Ratcliffe\\
    \small\it Dip.\ di Fisica, Univ.\ di Milano\\
    \small\it via G. Celoria 16, 20133 Milano, Italy}
\maketitle
\begin{abstract}%
We discuss the present situation with regard to polarised nucleon structure
function measurements. In particular we examine the status of the Bjorken sum
rule in the light of the recent data on the spin structure functions of (i) the
deuteron obtained by the SMC group at CERN and (ii) the neutron by the E142
group at SLAC. In order to fully exploit all the available data it is necessary
to study the complete Bjorken system of equations, which may be done in any of
several equivalent ways. Combining the new data with that already obtained for
the proton by the EMC group and earlier SLAC/YALE collaborations, together with
bounds obtained on the strange quark polarisation, we show that the Bjorken
system of equations is violated at about the $2\sigma$ level. We also show,
using unpolarised data, that arguments based on possible higher-twist
contributions are unable to account for this discrepancy. In conclusion, a
simple explanation in terms of a non-perturbative renormalisation of the
relevant Wilson coefficients is presented.
\\ [6pt]
\end{abstract}

\section{Introduction}
It has long been held (albeit by a limited community of ``spin physicists'')
that polarised \DIS\ can provide very stringent tests of factorisation in
\PQCD\ and the \QPM. Indeed, polarisation effects in general provide valuable
insight into the dynamics of hadronic interactions and are extremely sensitive
to the bound-state structure, so elusive to theoretical
approaches~\cite{PGR89}.
In particular, the \BSR~\cite{Bj66,Bj70} is a measurable quantity that can be
used for comparison between experimental data and theoretical predictions. The
experimental precision attainable now surpasses the ten-percent level while, on
the theoretical side, all relevant PQCD calculations have been carried out to
at least two-loop order (i.e., one-percent level) and for the \BSR\ itself to
three loops~\cite{Larin91}. Thus, one can consider such comparisons as serious,
indeed obligatory, tests of the applicability of PQCD to such processes.

Like many other sum rules, the \BSR\ is a direct consequence of the
operator-product expansion, as applied to \DIS\ and justified by asymptotic
freedom in \PQCD. In simple terms, all \DIS\ structure-function sum rules may
be derived from the hypothesized point-like behaviour of strong interactions at
short distances. However, only the Adler sum rule~\cite{Adler66} is directly
associated with a conserved hadronic current and is therefore unique in not
receiving higher-order \PQCD\ corrections, i.e., the Wilson coefficient
associated with the relevant light-cone operator is unity to all orders in (and
is independent of) perturbation theory. That is, the validity of the Adler sum
rule is not actually dependent on the validity of \PQCD, being founded on
symmetry principles of a deeper level, and its violation is virtually
inconceivable in the present field-theoretic framework. In contrast, the Wilson
coefficients associated with both the \BSR\ and Gross-Llewellyn Smith sum
rule~\cite{Gross69} are $1{-}\frac{\alpha_S}{\pi}$ (to first or next-to-leading
logarithmic order); consequently, both sum rules are inextricably linked to
perturbation theory and any discrepancy between their prediction and
measurement should be taken as evidence for a failure of \PQCD\ in predicting
the normalisation of the Wilson coefficients.

\section{Polarised Deep-Inelastic Scattering}

By performing \DIS\ with both beam and target polarised longitudinally, access
is gained to the structure function $g_1(x,Q^2)$~\cite{Hughes93}. In the \QPM\
this structure function has a simple relation to polarised quark distributions,
analogous to that of $F_1(x,Q^2)$:
\begin{equation}
\begin{array}{r@{\;=\;}l}
 g_1(x,Q^2) & \mbox{$\frac{1}{2}$} \sum_f e_f^2 \, \Delta q_f(x,Q^2) , \\
 F_1(x,Q^2) & \mbox{$\frac{1}{2}$} \sum_f e_f^2 \,        q_f(x,Q^2) ,
\end{array}
\end{equation}
where the sum runs over quarks and antiquarks of flavour $f$, $e_f$ is the
$f$-quark fractional charge and $x$ is the usual Bjorken scaling variable or
momentum fraction of the struck quark. The quark densities are defined in the
following manner:
\begin{equation}
\begin{array}{r@{\;=\;}l}
 \Delta q_f(x,Q^2) & q_f^+(x,Q^2) - q_f^-(x,Q^2) , \\
        q_f(x,Q^2) & q_f^+(x,Q^2) + q_f^-(x,Q^2) ,
\end{array}
\end{equation}
where $q_f^\pm(x,Q^2)$ are the densities of quarks of flavour $f$ and
positive or negative helicity with respect to that of the parent hadron.

Of course, experimentally it is an asymmetry that is actually measured: namely,
the ratio of the difference and sum of cross-sections for opposite helicity
configurations. The polarised structure function is then extracted, using the
following working definition:
\begin{equation}
 g_1(x,Q^2) = \frac {A_1(x,Q^2) \, F_2(x,Q^2)} {2x\,(1+R(x,Q^2))} ,
\end{equation}
where $R_1(x,Q^2)$ is the usual ratio of longitudinal to transverse unpolarised
structure functions and $A_1(x,Q^2)$ is the measured asymmetry.

Sum rules equate integrals over $x_B$ of such structure functions to
independently known quantities. For example, using SU(2) current algebra and
assuming scale invariance, Bjorken showed~\cite{Bj66,Bj70} that the
proton-neutron difference for $g_1$ was given by the axial-vector $\beta$-decay
constant of the neutron, $g_A$:
\begin{equation}
 \Gamma_1^{p-n} = \int_0^1 dx\,g_1^{p-n}(x,Q^2) = \mbox{$\frac{1}{6}$}\,g_A .
\end{equation}
Note that here there is apparently no longer any dependence on the energy
scale, $Q^2$. In fact, in \PQCD\ there are radiative corrections to the
right-hand side that depend on the running coupling constant $\alpha_S(Q^2)$
and thus only a logarithmic variation of the (experimental) left-hand side is
expected at most. This sum rule therefore becomes a rigorous prediction of
\PQCD, through the justification of asymptotic freedom that leads to an
approximate scaling behaviour.

\section{The Bjorken System of Equations}

The \BSR, first dismissed as worthless~\cite{Bj66} but later
revalued~\cite{Bj70}, is nevertheless difficult to test; it requires a
precision measurement of $g_1$ for both the proton and neutron over a sizeable
range of $x_B$ (roughly speaking, a coverage of $0.01{<}x{<}0.6$ is necessary).
While precise data have been available for the former target since
1988~\cite{EMC88,EMC89}, the first ever data for the latter have only recently
been published~\cite{SMC93,Lowe93,E142-93,Petr93}. Moreover, the full SU(3)
algebra of the baryon octet actually admits three independent quantities,
which, while having their natural expression in terms of up, down and strange
quarks, are better expressed in terms of the SU(3) axial-vector couplings:
\begin{equation} \label{BJsystem}
\begin{array}{c@{\;=\;}l}
 \PME{\AQO{u} - \AQO{d}}             &        g_A , \\
 \PME{\AQO{u} + \AQO{d} - 2\AQO{s}}  & \tilde g_A , \\
 \PME{\AQO{u} + \AQO{d} + \AQO{s}}   &        g_0 .
\end{array}
\end{equation}
The right-hand sides of the first two equations above correspond to known
constants ($g_A{=}1.2573{\pm}0.0028$~\cite{PDG92} and
$\tilde{g}_A{=}0.629{\pm}0.039$~\cite{PGR90}), but the third ($g_0$),
corresponding to the flavour-singlet axial-vector current, is unknown. Thus an
immediate prediction for, say, just the proton integral is not possible.

Let us remark in passing that a further combination of the $u$, $d$ and $s$
axial-current matrix elements is, in fact, accessible in $\nu$-$p$ elastic
scattering~\cite{Ahre87} and this would in principle allow an exact prediction
for single nucleon targets. Unfortunately, the precision of such measurements
is still too poor to permit any strong statement.

Good arguments can be made, however, for setting the strange-quark matrix
element equal to zero~\cite{Ellis74}: there are very few strange quarks in the
proton and they are concentrated below $x_B{\simeq}0.1$, where all correlations
are expected to have died out. Therefore the second two matrix elements of
eq.~(\ref{BJsystem}) were expected to be equal, thus leaving only two
independent quantities. This allows predictions to be made for the proton and
neutron separately, which can now be expressed in terms of the two axial-vector
constants and the strange-quark matrix element as
\begin{equation}
 \Gamma_1^{p(n)} = (-)\mbox{$\frac{1}{12}$}        g_A
                    + \mbox{$\frac{5}{36}$} \tilde g_A
                    + \mbox{$\frac{1}{3}$}  \PME{\AQO{s}} ,
\end{equation}
For clarity, the \PQCD\ corrections have been suppressed in the above, for more
complete expressions see, e.g., ref.~\citenum{Prep93a}. Conversely, these
equations may be used to extract the value of the strange-quark matrix element
given the value of $\Gamma_1$ for either nucleon, or (as in the case of the
SMC) the deuteron.

\section{The Data --- New and Old}

We shall now compare the experimental results obtained by the three experiments
with the theoretical predictions based on the Ellis-Jaffe sum
rule~\cite{Ellis74}:
\begin{eqnarray}
\makebox[2cm][l]{EMC}  \Gamma_1^p &=& \M 0.126 \pm 0.010 \pm 0.015 \nonumber \\
\makebox[2cm][l]{SMC}  \Gamma_1^d &=& \M 0.023 \pm 0.020 \pm 0.015           \\
\makebox[2cm][l]{E142} \Gamma_1^n &=&  - 0.022 \pm 0.006 \pm 0.009 \nonumber \\
[6pt]
%\end{eqnarray}
%\begin{eqnarray}
 \Gamma_1^p &=& \M 0.176 \pm 0.006 + \mbox{$\frac13\int$}\Delta{s} \nonumber \\
\makebox[0cm][l]{\raisebox{ 8pt}[0pt][0pt]{Ellis-}}%
\makebox[2cm][l]{\raisebox{-8pt}[0pt][0pt]{Jaffe }}
 \Gamma_1^d &=& \M 0.080 \pm 0.006 + \mbox{$\frac13\int$}\Delta{s}           \\
 \Gamma_1^n &=&  - 0.016 \pm 0.006 + \mbox{$\frac13\int$}\Delta{s} \nonumber
\end{eqnarray}
The short-fall in the EMC proton measurement with respect to the Ellis-Jaffe
prediction (taking ${\int}\Delta{s}{=}0$) is immediately obvious. This
observation led to the coining of the phrase {\em Spin Crisis\/}. A similar
(though less striking) observation may be made for the SMC deuteron integral.
In contrast, the neutron sum rule appears well satisfied by the E142 data. In
terms of the strange-quark contribution, both the EMC and SMC measurements
imply ${\int}\Delta{s}{\simeq}{-}0.15$ while that of E142 leads to
${\int}\Delta{s}{\simeq}{-}0.02$. Thus an important question is immediately
raised: ``How big can ${\int}\Delta{s}$ be?'' This is the subject of the next
section.

Before moving on, let us make a few remarks on the SMC data: unfortunately, the
errors are so large that they have little impact on any analysis although,
taken at face value, they appear to agree with the \BSR\ when combined with the
EMC data. However, a certain neglect of these data can be motivated as follows.
The particularly low value of the SMC deuteron integral is due, in roughly
equal measure, to the large negative values reported at low $x$ and to the very
early drop at large values of $x$. While the low-$x$ data are (barely)
compatible with the smooth extrapolation to zero of the E142 data, the converse
is certainly not true. Nevertheless, attempts of doubtful validity at combining
the two data sets data have been made; the result is agreement with the \BSR\
and (surprisingly) larger errors than the E142 data alone. This last fact
clearly underlines the lack of statistical correctness of such a procedure.
With regard to the high-$x$ region, we have shown~\cite{Prep93b} that the SMC
data violate a rigorous bound that may be constructed from the combined EMC and
unpolarised $F_2^{p,n}$ data; the spin data is used to eliminate the $u$-quark
spin density and the $F_2^{p,n}$ data then bounds that of the remaining
$d$-quark via positivity. Much of this discussion will be clarified by the
planned E142 neutron measurements down to $x{\simeq}0.01$, using a higher beam
energy.

\section{Bounding the Strange-Quark Spin}

Positivity trivially implies $|{\int}\Delta{s}(x)|{\le}s(x)$, although we
stress that equality would imply 100\% polarisation. Now, Regge phenomenology
tells us that the dominant contribution of the sea is is that of the pomeron
(behaving as $x^{-1}$), which we recall is spin independent. We have used the
very precise data on the unpolarised strange-quark distributions~\cite{CCFR92}
to bound the non-diffractive contribution and thus too the strange-quark
polarisation~\cite{Prep88,Prep90a,Prep91a}. The result of this analysis is the
following bound:
\begin{equation}
 \left| \int \Delta s \right| \le 0.021 \pm 0.001,
\end{equation}
which has been challenged (e.g., see refs.~\citenum{Brod88,Ioffe90}) with
various proposed constructions and/or parametrisations.
However, we have shown in the papers cited above that all such proposals fail
to agree with present experimental and phenomenological knowledge.

A quantitative measure of the discrepancy between the data and theory may be
obtained by performing a one-parameter fit for the strange-quark polarisation
to the data, with and without the bound, the results~\cite{Prep93a} are
summarized in table~\ref{fits}.

\begin{table}[hbt]
\caption{Results of a one-parameter fit for the strange-quark polarisation to
the data, with and without the bound.}
\setlength{\doublerulesep}{0.1pt}
\[ \begin{array}{|||l|||c|c|c|||} \hline\hline\hline
 \mbox{Expt.~Info.}  & \PME{\AQO{s}} & \chi^2/\mbox{DoF} & \mbox{C.L.} \\
\hline\hline\hline
 \mbox{EMC+Bound}          & -0.027 \pm 0.019 & \Z 6.8/1 & 0.9\% \\
\hline
 \mbox{EMC+SMC+E142+Bound} & -0.032 \pm 0.017 &   10.6/3 & 1.4\% \\
\hline
 \mbox{EMC+SMC+E142}       & -0.088 \pm 0.020 & \Z 6.3/2 & 4.3\% \\
\hline\hline\hline \end{array} \]
\label{fits} \end{table}

\section{Higher Twist and Higher-Order PQCD}

Not to be forgotten, of course, are the corrections to the various sum rules;
as already mentioned above, there are \PQCD\ higher-order corrections and the
possibility of higher-twist contributions (especially in the case of the E142
data) should also be considered. For the non-singlet currents the \PQCD\
corrections are known up to order $\alpha_S^3$~\cite{Larin91}. However, since
the value of $\alpha_S$ extracted experimentally is only valid up to order
$\alpha_S^2$ we only perform the analysis to this order. Although the inclusion
of the second-order contribution does indeed shift the prediction for the \BSR\
closer to the measured value, the effect is only of the order of a few percent.

The situation with regard to higher-twist contributions requires a little more
care. Ellis and Karliner~\cite{Ellis93} have claimed that according to QCD
sum-rule calculations~\cite{Bali90} the higher-twist contributions to the \BSR\
actually bring the prediction perfectly into line with the measured value. In a
similar analysis, Close and Roberts~\cite{Close93} show that extremely large
higher-twist contributions are required to explain all the data simultaneously.
The argument is as follows: the neutron data are taken at an average $Q^2$ of
only $2\,\mbox{GeV}^2$ while the EMC data taken at an average $Q^2$ of
$11\,\mbox{GeV}^2$; the problem then is to combine them at some common value of
$Q^2$. Experimentally it has been noted by each of the groups that the
asymmetry is independent of $Q^2$, within experimental accuracy. Thus it is
reasonable to use the very precise parametrisations of $F_2(x,Q^2)$ to extract
$g_1(x,Q^2)$ for any desired energy scale. Ellis and Karliner then propose to
take the EMC data and ``evolve'' them down to the scale of the E142 data; it
turns out that the variation in the integral of $g_1^p$ is very small. At this
point it is necessary to introduce a theoretical input, in the form of the
above-mentioned predictions for the higher-twist component of the \BSR. Such
predictions are clearly model dependent and the associated errors are typically
of the order of 100\%. Moreover, a recent paper~\cite{Ji93} has brought to
light an error in the input for these calculations, suggesting that the
contribution should be rather smaller and of the opposite sign.

\begin{figure}[hbt]
\centering
\epsfysize=10cm
\mbox{\hspace{0.5cm}}\epsfbox{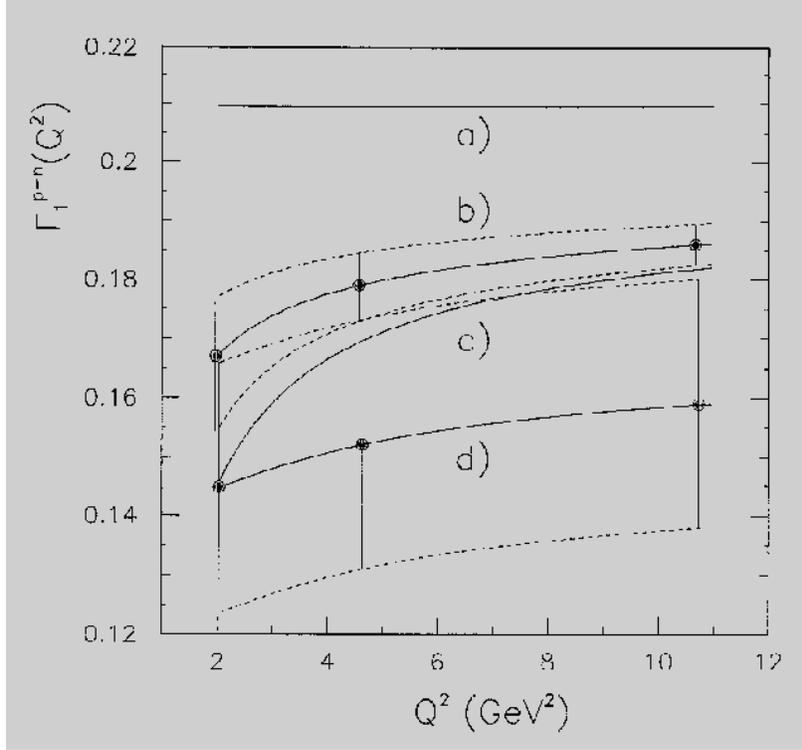}
\vspace{12pt}
\caption{The solid curves are the Bjorken sum rule: a) asymptotic value, b)
with the PQCD correction, c) PQCD and higher twist, d) experimental evaluation;
the errors for a) and c) are given by the dashed lines.}
\label{BJ} \end{figure}

\begin{figure}[hbt]
\centering
\epsfysize=10cm
\mbox{\hspace{0.5cm}}\epsfbox{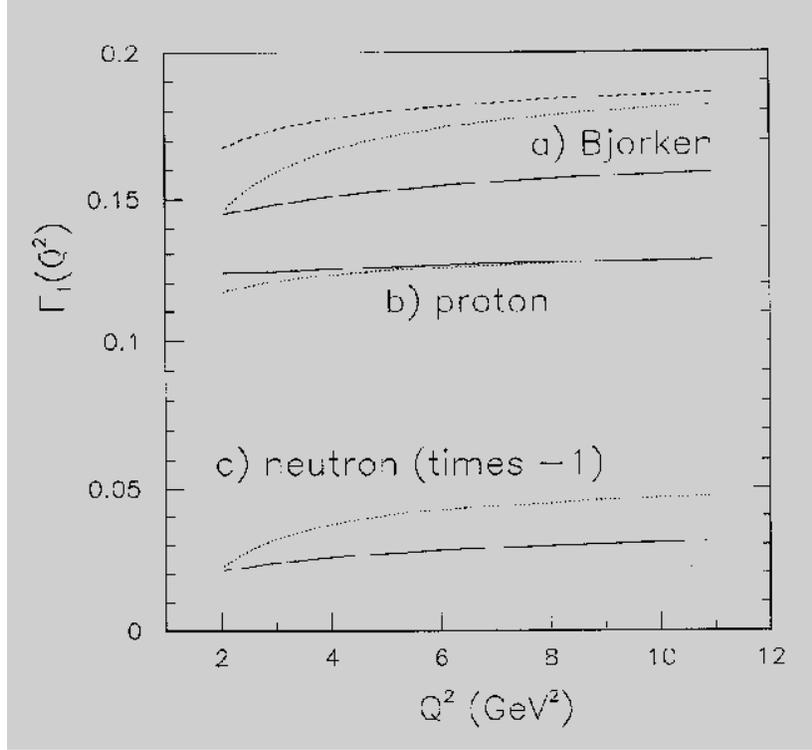}
\vspace{12pt}
\caption{The experimentally evaluated sum rules, as functions of $Q^2$, of a)
Bjorken, b) $\int_0^1g_1^p$ and c) $-\int_0^1g_1^n$ (solid curves) together
with the PQCD (dashed curves) and PQCD plus higher-twist predictions (dotted
curves).}
\label{BJetc} \end{figure}

In any case, it is clearly preferable to try to avoid the the necessity of
higher-twist estimates; this can be achieved by simply evolving the E142 data
to the higher scales of the EMC experiment. It turns out, using this procedure,
that the variation in the integral of $g_1^n$ is also small, reflecting a
negligible higher-twist component. Thus the value of the \BSR\ at
$Q^2{=}11\,\mbox{GeV}^2$ is only slightly affected by this procedure; the
integrated values for the proton, neutron and \BSR\ at $Q^2{=}11\,\mbox{GeV}^2$
are
\begin{equation}
\begin{array}{ll@{\;=\;}c@{\,\pm\,}c}
\mbox{EMC}                      & \Gamma_1^p     & \M 0.128 & 0.019 \\
\mbox{E142}                     & \Gamma_1^n     &  - 0.032 & 0.011 \\
\makebox[3cm][l]{\BSR~(expt.)}  & \Gamma_1^{p-n} & \M 0.160 & 0.022
\end{array}
\end{equation}
to be compared with
\begin{equation}
\begin{array}{ll@{\;=\;}c@{\,\pm\,}c}
\makebox[3cm][l]{\BSR~(th.)}    & \Gamma_1^{p-n} & \M 0.192 & 0.003
\end{array}
\end{equation}
Our results for the $Q^2$ dependence of the various sum rules are displayed
graphically in figs.~\ref{BJ} and \ref{BJetc}. In table~\ref{tab:HT} we show
the result of separating the various leading- and higher-twist contributions to
the sum rules via a knowledge of the $Q^2$ dependence of the unpolarised
structure functions.

\begin{table}[hbt]
\caption{The proton and neutron integrals and higher-twist corrections, the
fourth and fifth columns give the two corrections and the last column contains
the corrected (i.e., leading-twist) integrals.  Note that rounding errors cause
slight discrepancies between the various figures.}
\setlength{\doublerulesep}{0.1pt}
\[ \begin{array}{|||r|c|r|r|r|r|||} \hline\hline\hline
 Q^2 & \mbox{Target} & \Gamma_1^{\rm expt.} &
 \delta\Gamma_1^{\rm DHT} & \delta\Gamma_1^{\rm TM} &
 \Gamma_1^{\rm LT.}\qquad \\ \hline\hline\hline
      &  n  & -0.020 & -0.003 &  0.000 & -0.023\pm0.010 \\ \cline{2-6}
  2.0 &  p  &  0.124 & -0.005 & -0.003 &  0.116\pm0.017 \\ \cline{2-6}
      & p-n &  0.144 & -0.002 & -0.002 &  0.140\pm0.020 \\ \hline\hline\hline
      &  n  & -0.026 & -0.001 &  0.000 & -0.028\pm0.010 \\ \cline{2-6}
  4.6 &  p  &  0.125 & -0.001 & -0.001 &  0.123\pm0.017 \\ \cline{2-6}
      & p-n &  0.152 &  0.000 & -0.001 &  0.151\pm0.020 \\ \hline\hline\hline
      &  n  & -0.031 & -0.001 &  0.000 & -0.032\pm0.010 \\ \cline{2-6}
 10.7 &  p  &  0.128 &  0.000 &  0.000 &  0.127\pm0.017 \\ \cline{2-6}
      & p-n &  0.159 &  0.000 &  0.000 &  0.159\pm0.020 \\ \hline\hline\hline
\end{array} \]
\label{tab:HT} \end{table}

\section{Extracting Spin Densities from the Data}

Adopting the attitude that the overall normalisation of the proton matrix
elements of the light-cone current operators is unreliably estimated in \PQCD,
it might be asked if there is any way to extract the three unknowns (i.e.,
$\Delta{u}$, $\Delta{d}$ and $\Delta{s}$) from essentially only two independent
data sets (proton and neutron). Now, while overall normalisations are often
unreliable, ratios are well determined in the standard broken SU(6) picture of
the nucleon; thus let us take as a reliable quantity only the ratio
\begin{equation}
 \frac{\Delta{d_v}}{\Delta{u_v}}
 = - \frac{(D-F)}{2F} \simeq - \frac{1}{3},
\end{equation}
where $F$ and $D$ are the usual baryon-octet $\beta$-decay constants; in SU(6)
the ratio $\frac{F}{D}{=}\frac{2}{3}$, experimentally it is
${\simeq}0.6$~\cite{PGR90}. Using this relation to substitute for $\Delta{d_v}$
in the expressions of for $\Gamma_1^{p,n}$ and eliminating $\Delta{u_v}$
between the two, we arrive at the following relation:
\begin{equation}
 \Gamma_1^n = -\mbox{$\frac{1}{11}$}\Gamma_1^p+\mbox{$\frac{2}{3}$}\Delta{s} .
\end{equation}
In obtaining the above, we have also included the phenomenological factor-two
suppression of the strange over non-strange sea (although the final numerical
results are rather insensitive to the actual value used).
Thus, from our bound on $\Delta{s}$ and the EMC value for $\Gamma_1^p$, we
could have predicted
\begin{equation}
 0.002 \ge \Gamma_1^n \ge -0.026 ,
\end{equation}
in rather good agreement with the SLAC data.

\begin{figure}[hbt]
\centering
\epsfysize=10cm
\mbox{\hspace{0.5cm}}\epsfbox{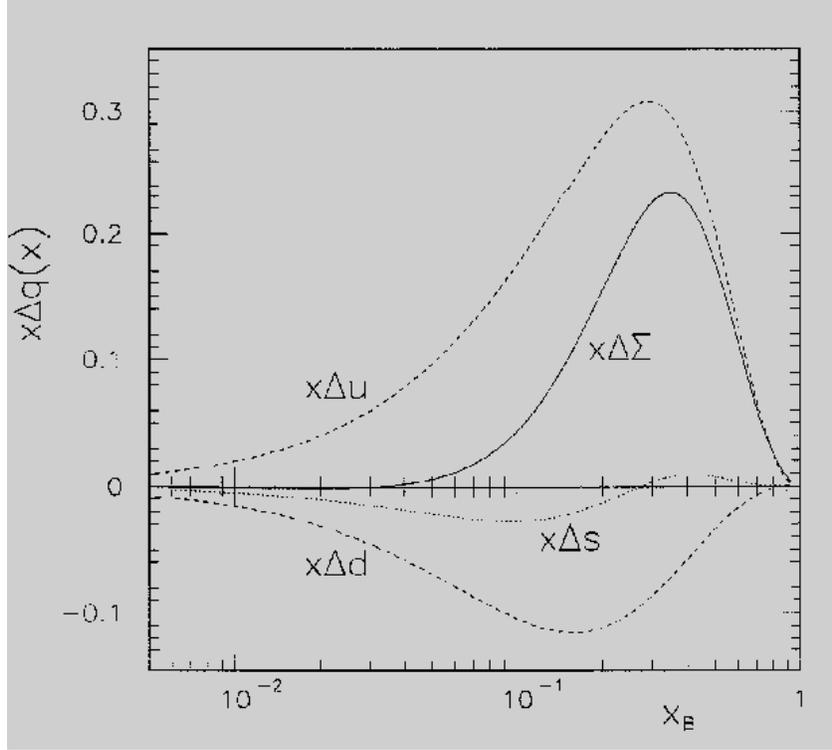}
\vspace{12pt}
\caption{Our extraction of the quark spin distributions: the upper and lower
dashed lines are the $u$ and $d$ quarks respectively, the dotted line is the
$s$ quark and the solid line is the resulting total quark contribution.}
\label{ourfit} \end{figure}

Alternatively, we can turn these expressions around and extract the various
integrated quark spin-dependent densities, using both EMC and SLAC data:
\begin{equation}
\begin{array}{r@{\;=\;}l@{\hspace{22mm}}r@{\;=\;}l}
 \Delta{u_v} & \M 0.72 & \Delta{u}      & \M 0.66 \\
 \Delta{d_v} &  - 0.24 & \Delta{d}      &  - 0.30 \\
 \Delta{u_s}=\Delta{d_s} &  - 0.06 & \Delta{s}  &  - 0.03 \\ [6pt]
 \multicolumn{3}{r@{\;=\;}}{\Delta{\Sigma}{\;=\;}
    \Delta{u}{+}\Delta{d}{+}\Delta{s}}  & \M 0.33
\end{array}
\end{equation}
(the errors on the above quantities are typically of the order of $0.02-0.03$).
Notice (i) the strange-quark component is entirely compatible with the bound
and (ii) the total quark spin, $\Delta{\Sigma}$, is a sizeable fraction of that
of the proton. With the additional standard assumption of an extra factor of
$1{-}x$ for the valence $d$ density with respect to $u$, the distributions as
functions of $x$ may be obtained; the corresponding curves are shown in
fig.~\ref{ourfit}.

\section{Conclusions}

Let us preface the closing remarks by noting that all of the above is
crucially dependent on the validity of the original EMC data, which we have
taken at face value. Thus, given the enormity of the conclusions to which we
are ineluctably drawn, it is important to stress the urgency of remeasuring the
proton spin sum rule to the highest possible precision.

Although the \BSR\ only fails to be satisfied experimentally at something like
a $1\frac{1}{2}\sigma$ level, the complete system of equations including the
strange-quark polarisation bound is violated at a level of
$2\frac{1}{2}\sigma$. We have also shown that a mild assumption of incorrect
evaluation of the overall normalisation for hadronic matrix elements in \PQCD\
leads to a picture of the nucleon in which a large fraction of the spin is
carried by the quarks and the strange-quark spin is very small, i.e., precisely
the picture existing before the beginning of the {\em Spin Crisis\/}.

To conclude, let us just recall that a correct prediction of the proton spin
structure function had appeared in the literature~\cite{Gian85} in 1985, well
before the EMC results. Implicit in the ACD Fire-String model that led to this
prediction, is the almost vanishing neutron spin asymmetry, as corroborated by
the E142 results. The model, with an asymmetry driven essentially by a broken
SU(6) symmetry, has a final-state structure in which non-physical (i.e.,
coloured) states are explicitly barred from contributing. The two contributions
to \DIS\ are (i) the single fire-string diagram, where quark helicity
conservation makes itself strongly felt and the broken SU(6) structure leads to
essentially vanishing neutron asymmetry; and (ii) the double fire-string
diagram where the effects of quark polarisation are so dilute as to be
negligible in all cases. The differences between the explicitly physical
final-state structure of ACD and that implicitly unphysical in \PQCD\ explain
the failure of the latter to obtain the correct normalisation (or Wilson
coefficients) of the nucleon matrix elements in question.

\section{Note added}

Since presenting this talk the next-to-leading PQCD corrections to the
Ellis-Jaffe sum rule (in other words for the flavour-singlet piece) have been
made available~\cite{Larin94}. Thus, given the value of any single nucleon
integral at some $Q^2$, one can perform the two-loop evolution to any desired
$Q^2$ scale. We have checked the effect of this approach and find negligible
difference with the results presented above, although it should be remarked
that the theory-data discrepancy is then, in fact, slightly accentuated.

%=============================== END OF TEXT =================================%

% \begin{references}

% \end{references}

\end{document}